\documentclass{ws-procs975x65}

\begin{document}

{\hfill RESCEU-7/10}

{\hfill UNF-Astro-3-1-10B}

{\hfill \ }

\title{DO EVAPORATING 4D BLACK HOLES FORM PHOTOSPHERES AND/OR CHROMOSPHERES?}

\author{J. H. MACGIBBON}

\address{Department of Physics, University of North Florida,\\
Jacksonville, Florida 32224, United States\\
$^*$E-mail: jmacgibb@unf.edu}

\author{B. J. CARR}

\address{Astronomy Unit, Queen Mary, University of London, Mile End Road,\\
London E1 4NS, United Kingdom\\
Research Center for the Early Universe, University of Tokyo, Tokyo 113-0033, Japan\\ 
E-mail: B.J.Carr@qmul.ac.uk}

\author{D. N. PAGE}

\address{Department of Physics, University of Alberta,\\
Edmonton, Alberta T6G 2G7, Canada\\
E-mail: don@phys.ualberta.ca}

\begin{abstract}
Several authors have claimed that the observable Hawking emission from a microscopic black hole is significantly modified by the formation of a photosphere or chromosphere around the black hole due to QED or QCD interactions between the emitted particles. Analyzing these models we identify a number of physical and geometrical effects which invalidate them. In all cases, we find that the observational signatures of a cosmic or Galactic background of black holes or an individual black hole remain essentially those of the standard Hawking model, with little change to the detection probability.
\end{abstract}

\keywords{Primordial black holes; Hawking radiation; black hole photospheres.}

\bodymatter

\section{Introduction}\label{aba:sec1}
Microscopic black holes are of great interest in astrophysics and cosmology\cite{C05}. A 4D  black hole with a Hawking temperature~\cite{H} $T_{bh}= 1.06\ {\rm GeV}/\left( M_{bh}/10^{13}\ {\rm g} \right)$ should emit all available particle species which appear non-composite compared with the wavelength of the radiated energy. Once $T_{bh}\gtrsim \Lambda_{QCD}\simeq 200 - 300$ MeV, quarks and gluons should be directly emitted\cite{MW} and then decay into stable species. It has been claimed that interactions between emitted particles significantly modify the astrophysically observable spectra from such black holes. Most scenarios are based on the Heckler\cite{HE} model in which two-body bremsstrahlung and pair-production interactions form a QED photosphere at $T_{bh}\gtrsim 45$ GeV and a QCD chromosphere at $T_{bh}\gtrsim \Lambda_{QCD}$. Here we summarize our recent detailed analysis\cite{MCP,PCM} of interaction models and discuss various points which prevent the development of photospheres and chromospheres. 

\section{Is the Heckler Model Correct?}
The Heckler model takes the two-body bremsstrahlung (and corresponding pair-production) cross-section\cite{Haug} to be $\sigma _{brem} \approx \left( 8\alpha ^3 / {m_e} ^2 \right) {\rm ln}\left(2E / m_e \right)$ where $E$ is the energy of the initial electrons ($e^\pm$) in the center-of-momentum (CM) frame and $m_e$ is the electron mass. In the CM frame, the average momentum exchanged is $\sim m_e$ (implying particles must be within $\sim 1/m_e$ of each other to interact), the average angle between the final (`on-shell') electron and outgoing photon is $\phi_{av} \sim m_e/2E$, and the average energies of the final electron and photon is $\sim E/2$. Even if a photosphere develops, most particles continue to move radially away from the black hole. Since the average energy\cite{P3} of the Hawking emitted electrons is $E_{peak}\simeq 4T_{bh}$ and $\phi_{av}\sim m_e/2E$, a particle emitted by a $T_{bh}\sim 1$ GeV black hole would have to undergo $\mathcal{N}\sim 10^6$ scatterings to deviate $0.1$ radians from the radial direction. The form of $\sigma_{brem}$ is accurate but, as discussed below, its application must be modified in the black hole context. The following corrections to the Heckler model must be applied:

(i) The black hole rest frame can be shown\cite{MCP} to be the CM frame for most pairs of emitted particles but two particles moving in a similar direction, whose CM frame is highly Lorentz-boosted relative to the black hole, will not interact near the black hole. Thus each particle is surrounded by an `exclusion cone' of particles with which it can not interact significantly. Once the particle is a distance $d$ from the black hole, the transverse distance to the nearest one with which it can interact is $x_T\sim d$. However, from above we require $x_T\lesssim 1/m_e$. Thus particles must be within $\sim 1/m_e$ of the black hole to interact. This point was omitted in the Heckler model.

(ii) Because the black hole flux radiates outward from a central point, the mean free path formula $\lambda \approx \left(n\sigma_{brem}v_{rel}\right)^{-1}$ employed in the Heckler model is not appropriate and must be replaced by a radial description. Additionally, particles created at the black hole do not originate at an infinite distance from their interaction region, contrary to assumptions in the $\sigma_{brem}$ derivation. Thus the effective interaction cross-section for particles emitted by the black hole must be less than $\sigma_{brem}$.

(iii) Any two particles must be in causal contact to interact. For two particles within $\sim 1/m_e$ of the black hole, it can be shown\cite{MCP} that this requires the time between their Hawking emission, $\Delta t_e$, to be less than the causal interval $\Delta t_c \sim 1/(\gamma m_e)$ where $\gamma \sim E_{peak}/m_e$. However, for the Hawking flux $\Delta t_e\sim 200/E_{peak} > > \Delta t_c$. Thus a negligible fraction of emitted particles are causally connected within $\sim 1/m_e$ of the black hole. This point was omitted in the Heckler model.

(iv) The Heckler model assumes that the distance required for the formation of a final on-shell electron and its bremsstrahlung photon is $d_{form}\sim 1/m_e$ in the CM frame. Applying the Heisenberg Uncertainty Principle to outgoing particles separated by $\phi_{av}\sim m_e/2E$, the correct distance is $d_{form}\sim E/{m_e}^2$ in the CM frame. Thus any multiple interactions experienced by an electron within $\sim 1/m_e$ of the black hole are off-shell interactions and so are strongly suppressed by the LPM effect\cite{LPM}. This was omitted in the Heckler model.

We conclude that the causal contact requirement and the suppresion of multiple scatterings prevent the development of a Heckler QED photosphere. Analoguous arguments also prevent a QCD photosphere developing when $T_{bh}>>\Lambda_{QCD}$. 

\section{QCD Chromosphere when $T_{bh}\sim \Lambda_{QCD}$?}
When $T_{bh}\sim \Lambda_{QCD}$ (or more strictly, $E_{peak}\simeq \Lambda_{QCD}$), the Hawking flux is damped by the grey-body factors which apply near rest mass thresholds\cite{P3,MW}. This increases $\Delta t_e$ and strengthens the causality constraint. Additionally, the limited available energy per emitted particle ($E\sim \Lambda_{QCD}$) and particle physics conservation laws severely constrain the number of final particles that can be produced per emitted particle. In particular, gluon bremsstrahlung is insignificant when $T_{bh}\sim \Lambda_{QCD}$ because the lowest colourless state is a pion and $m_\pi \sim 140$ MeV. Therefore, the black hole situation is not analogous to the $200$ GeV per nucleon, gluon-saturated, high baryon/antibaryon asymmetry quark-gluon plasma seen at RHIC. Instead, it resembles the conditions in $e^+ e^-$ accelerator events, for which the smooth transition around $\Lambda_{QCD}$ from the direct pion regime to the quark/gluon-mediated regime sets in when the pions (and hence the constituent quarks) are relativistic. This implies that when the black hole goes from directly emitting pions to directly emitting quarks and gluons, the latter are relativistic and do not linger near the black hole. Hence a QCD photosphere should not develop when $T_{bh}\sim \Lambda_{QCD}$.

\section{Observational Consequences}
Further photosphere and chromosphere scenarios\cite{BY,Bug,DH,DC3,CL,DK3,Kap,MOS,R}\,, including those not based on the Heckler model, are analyzed in Refs.~\refcite{MCP} and \refcite{PCM}. In all cases, we find that the observational signatures and detection probabilities of an individual black hole or a cosmic or Galactic halo background of primordial black holes remain essentially those\cite{MC} of the standard Hawking model, although as first noted in Ref.~\refcite{PCM} inner bremsstrahlung photons generated by charged particles as they accelerate away from an individual black hole dominate over the direct Hawking photons at low photon frequencies. We cannot exclude, however, the possible enhancement of black hole bursts by unusual ambient fields or new physics. Whether a chromosphere can develop around a higher-dimensional black hole is also presently an open question.

\end{document}